IAC-22,E5,1,7,x72645

# Holistic Outpost Design for Lunar Lava Tubes

Anna L. Vock[a]*, Dr. Tommy Nilsson[b]

[a] *Bauhaus-University Weimar, Geschwister-Scholl-Straße 8/15, 99423 Weimar, Germany* , annalvock@gmail.com
[b] *European Space Agency (ESA), European Astronaut Centre (EAC), Linder Höhe, 51147 Cologne, Germany*
\* Corresponding Author

**Abstract**
As the space industry continues its rapid development, humanity is poised to expand beyond Low Earth Orbit (LEO), seeking to establish permanent presence on the Moon and beyond. While space travel has traditionally been the domain of a small number of highly specialized professionals, a new era of human exploration, involving non-space actors and stakeholders, is now becoming a reality.
In spite of this development, most space habitats are still designed for a narrow target group. This paper seeks to address this deficit by rethinking the established design approaches, typically limited to tackling technological and physiological challenges of human space exploration (such as radiation or hypogravity), by instead adopting an interdisciplinary 'big picture' perspective encompassing social, psychological and cultural aspects of future space habitats.
In collaboration with the European Space Agency (ESA), we have conducted an extensive literature review and interviews featuring both industry experts, as well as members of the public. On these grounds, we then present a broad overview of key considerations and relevant human factors surrounding the design of future large-scale space habitats with a lifetime that spans generations of settlers.
Drawing on elements from architecture, urban studies and design fiction, we synthesize our findings into a hypothetical lunar settlement concept. Realized by in-situ resource utilization (ISRU) and 3D printing, we propose the construction of sub-surface living spaces in lava tubes, with respect to the rough internal terrain, close to the lunar poles.
We take a holistic human-centered approach prioritizing the fostering of community, belonging and identification with the habitat. To this end, the concept encompasses 7 different module types each tailored to meet specific human needs such as private, social as well as multi-purpose spaces. This approach purposefully distances itself from traditional centrally planned settlements, by instead enabling inhabitants to autonomously and organically grow their colony over an extended period of time. The theoretical urban development is illustrated by a conceptual 25-years plan starting from 2030. Our aim is to present a concept that is generalizable and applicable to destinations beyond the Moon.
By elaborating and reflecting on our concept, this paper seeks to demonstrate the importance of a trans-disciplinary approach to designing thriving sustainable colonies beyond LEO. We demonstrate the potentially key role of design as mediator in advancing macro-strategies promoting thriving existence and sustainable growth. With this approach we tackle big-picture questions about humanity's future and prospects amongst the stars..
**Keywords:**

**Acronyms/Abbreviations**
ALM: Additive layer manufacturing
CAD: Computer-Aided Design
ESA: European Space Agency
F+P: Foster + Partner
FDM: Fused Deposition Modeling
HF: Human Factors
ISRU: In Situ Resource Utilization
ISS: International Space Station
LEO: Low Earth orbit
NASA: The National Aeronautics and Space Administration
SOM: Skidmore, Owings & Merrill Architects
TRL: Technology Readiness Level
VR: Virtual Reality
WWWWWH: Who, What, Where, When, Why, and How
XR: Extended Reality

1. **Introduction**

Humanity is once again returning to the Moon, this time to stay, as the Artemis mission slogan proclaims. 50 years after the last landing, our understanding of the Moon has changed drastically. Recent research indicates the presence of water and other resources crucial for sustaining human life on the Moon. Emboldened by these findings, space agencies around the world are now in the process of drawing up ambitious plans for establishing permanent lunar settlements in the near future.

Furthermore, an established base will be the launchpad into deep space for future generations. To establish this permanent presence, focus must be set on the habitability of our future homes. Historically, the development of space habitats, such as Skylab or the International Space Station (ISS), has focussed on survivability rather than liveability, let alone on how to sustain a community of space settlers in the long term. Against the backdrop of recent developments, it is now high time to challenge this approach by refocusing our lens on humanity's long-term and sustained future among the stars. This paper seeks to counter the established engineering mindset by proposing an alternative designerly approach to this all-encompassing challenge, utilizing a design thinking and human centered approach, comprising design methodologies in use today. The goal is to demonstrate the viability of this approach by envisioning a hypothetical settlement, giving shape to its structures using the possibilities opened up by current technological advancements (ISRU and 3D Printing). The idea is communicated by illustrations, renders and 3D models.

*1.2. Current landscape and research goal*

Up to today, habitability of space structures has been low. The reason behind this being the origin of development, which is rooted in military advancement, conceptualization by engineers and the use of developmental processes originally designed for machines and other non-human systems. As Mary Roach put it in her book *Packing for Mars*, "*To the rocket scientist, you are a problem. You are the most irritating piece of machinery he or she will ever have to deal with. You and your fluctuating metabolism, your puny memory, your frame that comes in a million different configurations. You are unpredictable. You're inconstant. You take weeks to fix. The engineer must worry about the water and oxygen and food you'll need in space, about how much extra fuel it will take to launch your shrimp cocktail and irradiated beef tacos. A solar cell or a thruster nozzle is stable and undemanding. It does not excrete or panic or fall in love with the mission commander. It has no ego. Its structural elements don't start to break down without gravity, and it works just fine without sleep.*" [1]. Human Factors (HF) have not received the necessary attention. Moreover, the definition of HF will undergo a change with the development of long-duration missions. While short duration missions can simply provide survivability, covering the most basic human needs, long term missions will need to address a multitude of partially unforeseeable factors linked to human well-being on multiple levels, therefore making the space habitable. "*Habitability is a general term which connotes a level of environmental acceptability. The requirements for conditions to be "habitable" change dramatically with circumstances. For brief periods, almost any arrangement that does not interfere with the health of the individuals or the performance of their jobs would be acceptable. Over the long term, conditions must support not only individuals' physical, but also their psychological health.*" [2].

Contemporary habitat designs, which are brought forward by non-governmental companies, or in some cases in partnership with space agencies, such as the European Space Agency (ESA), involve a more human-centered approach to habitat design than previous space habitats have. Nevertheless, a majority of them follow methodological patterns that prioritize technical feasibility over anything else. The process of establishing a successful settlement on the Moon has been suggested to unfold over three stages [3]. The first stage is exploratory in nature, focusing on mapping of the lunar landscape in search of a viable settlement location. The second stage involves the establishment of an initial outpost using resources and supplies delivered from Earth but constant presence of habitants. Finally, the third stage sees the settlement achieve a high degree of sustainability and self-sufficiency by exploiting available lunar resources.

With the exploratory stage well underway, we now know of several strategically viable regions, such as the Shackleton crater area on the lunar south pole (with its permanently shadowed floor) where one might locate water in the form of ice as well as other valuable substances [4]. The prospect of employing innovative use of in-Situ-resource-utilization (ISRU) in combination with 3D printing or regolith sintering techniques in such locations has indeed opened up new possibilities for long term habitation. Nevertheless, contemporary concepts of potential settlements are typified by a Master-Plan approach, with individual modules assembled and organized following a strictly predetermined arrangement, as opposed to taking a more organic bottom-up approach. Moreover, such design concepts are typically counting on prefabricated, pressurized habitation modules transported from Earth to the lunar surface, with little attention being given to the third settlement stage and the manner in which an outpost ought to grow and organize itself once reaching a degree of resource autonomy. This focus has resulted in crucial questions being neglected: What comes after, once we have established a preliminary base with permanent settlers from all sorts of cultural or educational backgrounds? How might we live in the future, together, as an interplanetary species?

In search of an answer, this paper proposes a novel habitat concept that explores the benefits of a design thinking approach, applied to a potentially viable environment inside pre-existing caves of lava tubes, which have been recently discovered on moon and mars. Specifically, the concept takes on the form of a regolith 3D printed village that is especially adapted to the human needs, such as community, belonging and identification with the habitat, which is absent in most contemporary concepts. In so doing, we outline a multidisciplinary approach, which is deemed necessary for future habitat designs, if lunar villages should become long term solutions to human space travel beyond LEO. Choosing a human centered design approach restructures the entire developmental process and provides a concept that is innately developed around the human need for a stable community in the inhabitable environment of the Moon.

The structure of the paper is as follows: To follow the design thinking process, we carried out a qualitative study centered around the collection of relevant considerations through interviews with industry specialists as well as people of the public. In addition, a review of relevant literature (section 2) was likewise conducted (section 3). Following this research phase, specifications of a probable future were defined (section 4). Using conventional design methods such as Persona and User-Journey-Mapping to better understand human needs in this particular environment, concepts were iteratively explored and developed (section 5) . Using social and architectural observations of towns from Christopher Alexander's *A Pattern Language* [5], the final design was constructed via computer-aided design (CAD) and visualized as renders to give shape to the concept (Section 5). Next to the innovative design approach using design methodology that can be adapted to future habitat development studies, this concept provides inspirational imagery of possible future habitats which should provide input to designers and engineers alike.

## 2. Related work

Studies on habitability of past future habitat designs have been relatively common at least since the early 2000's, which was also when some of the first studies of interdisciplinary work, by including design and architecture in early project development phases, for human space travel was developed. Driving observations and studies were made on this field by Dr. Imhof, Dr. Schlacht and Dr. Häuplik-Meusburger for example which are still relevant today. In the past, around the 1960-1970's, architecture studios like Archigram and Superstudio critically challenged the question of future architecture and life in space with the help of illustrations or collages. These often radical design ideas originate from a very uncertain time, shortly after World War II and at the beginning of the Cold War, which gave rise to a debate on the future of daily life and therefore architecture too. With their speculative design they criticized ways of life and the architecture of their time and with optimism looked into the future [6]. Archigram's "Plug-In-City", a  pod-living concept, shares some design similarities with the ISS. Both are based on a modular construction that can be adapted depending on the user's needs. In Archigam's concept though, the Plug-In-City grows over the entire planet, while the ISS can not be considered a city at all, one would argue. And while the studios drove the new idea of architecture without the architect, a democratizing stance, the ISS's configuration is never decided on by the inhabitants themselves. It is interesting to observe designers and architects challenging current states of certain systems and seeking to provide answers. As seen with the ISS however, they are seldom implemented to the full extent of the vision.

Moving forward to today, the following examples are architectural concepts that have received exceptional media attention in the recent past. After analyzing their components and the way in which they were designed, we'll move on to laying out the drawbacks and outline the existing gap that this work seeks to fill.

Foster + Partners developed a habitat concept in 2012 for 4 people, located on the south pole of the Moon. It involves a cylinder containing an inflatable module that expands into a protective habitable structure upon being deployed  by the Artemis mission's lunar lander . Layers of regolith are then shoveled over the inflated module by multiple robots and then synthered, forming a solid construct [7]. This concept is considered a pioneering project in the 3D printed lunar habitat domain. Main focus of this specific work was to assess the feasibility of 3D printing habitats in vacuum. The feasibility of 3D printing in earth-like conditions (including room temperature) has already been established through internal experiments during which F+P printed a 1,5 tonne building block to demonstrate the innovative design of the supporting structure of the habitat [8]. Even though the images produced and the knowledge of 3D printing possibilities on the lunar surface are visionary and inspirational to many projects that followed, the concept focused mainly on the technical aspects of 3D printing in space, rather than providing a habitat that is tailored to human needs.

Similarly, Liquifer Systems Group brought forward their RegoLight concept in 2015. The objective of this project was to advance the additive layer manufacturing (ALM) (or 3D printing capabilities) from a low technical readiness level (TRL) to a medium TRL, which was successfully fulfilled by the conclusion of the project after 2 years [9]. Their concept contains, similarly to the one of Foster + Partners, a combination of an inflatable interior shell that provides a pressurized environment for the inhabitants, covered by processed solar sintered regolith.

The most recent habitat concept was the *Lunar Village* presented by Skidmore, Owings & Merrill Architects (SOM) at the Biennale in Venice 2021, designed in collaboration with MIT Media Lab and ESA in 201 . Beyond its detailed design, it also introduces a new design language to the pre-existing habitat concepts. In contrast to the other concepts, the Lunar Village was developed by an international, interdisciplinary team of designers, architects, engineers and stakeholders. It consists of vertically arranged, semi-inflatable habitats envisioned to be transported to the lunar surface by adapted giant lunar landers. Great emphasis has been placed on the circular economy of the settlement, providing each habitat with its own life support systems and compact living and workspaces. In designing the interior, attention was paid to the configuration of spaces, lighting conditions, and high ceilings (due to reduced gravity) [10] which proves the consideration of human needs. Beyond that, the resulting paper published at the 2019 IAC [11] highlights the interdisciplinarity of the project, which is described as *necessary* for such a large project that spans an entire settlement and is not included in any other related work so far.

All the above mentioned concepts are based on a master-plan approach that enables modular expansion over longer periods of time, as well as easy replacement or extension due to the modularity. Apart from that, 2 out of the 3 focus on ISRU and ALM processes as a means to reduce cost and waste production through transport from Earth. Only one of the concepts shows a focus on the human experience of living in space (SOM's), while the others focus mainly on the feasibility of the chosen production technology.

The reason for this gap might lie in the current state of space exploration and the ongoing developments of long term lunar habitation solutions. Currently, the next habitation environment to follow the ISS is the Gateway, scheduled to be deployed in Moon's orbit by the late 2020's, therefore, there is no inherent need for surface habitat concepts. Taken from a personal conversation with an ESA specialist, there is simply no financial leeway to explore these distant possibilities currently, which is also why all of the mentioned works are from external developers and the studies cover mainly production (focussed on materials and construction), rather than the human-centered design side of habitat structures.

Surprisingly, none of the existing concepts incorporated the new information and therefore possibility of underground construction. Skylights, discovered through images of Selene in 2009, lead us to believe there are lava tubes underneath the surface that might be multiple kilometers long and up to multiple meters high [12]. This would provide an ideal permanent habitat site, due to the protection underneath the lunar surface, shielded from asteroid impact and the harsh temperature changes through the dark and light periods as well as radiation. The location does however bring many challenges that impact humans and material greatly, for example the constant darkness and the razor sharp regolith formations.

The design language employed by the majority of extra-terrestrial habitat concepts remains heavily restricted by the mono-disciplinary engineering-centric approach driven by the space industry, which is leading the developments and dialogues. In the remainder of this paper, we will attempt to challenge this.

### 3. Methodology

Although the goal of this project is very conceptual due to the temporal location and circumstances that are still far from our current state, (or perhaps precisely because of this), this work needs a clear methodological anchoring that, together with the creativity of the authors, brings to light a stimulating product. Most of the methods utilized come from the Delft Design Guide[13] and have been selected for the User-Centered-Design Approach. It is important to highlight this specific approach, as it is closely interconnected with the stated goals of this paper, namely the proposition of a lunar settlement that is conducive to the inherent needs of its residents. This process follows a divergent and convergent design approach, which makes up the design thinking method utilized in this project (Friis).

Design thinking is a problem solving approach developed by the Hasso-Plattner-Institute of Design (HPI D-School) at Stanford. It consists of 5 main stages: Empathize, Define, Ideate, Prototype and Test. In order to empathize with the users, we have to explore and familiarize ourselves with their needs. Following this stage, user needs ought to be defined, thus enabling the ideation phase. During this phase, creativity is used to develop concepts, which will then be prototyped and tested. This process is not linear but iterative. Stages can be running simultaneously or change order, too [14]. This stands in great contrast to the relatively linear development of Technical Readiness Levels used in engineering.

In order to provide an overview of current developments in space exploration, and more specifically in human-lead exploration of other celestial bodies such as the Moon, an in-depth literature review was conducted. In parallel, a set of 4 open interviews with relevant experts were carried out. These included:

- Antonio Fortunato, a specialist in ISS function and build
- Aidan Cowley, an ESA science advisor involved in previous habitat conceptualizations, including the aforementioned SOM's lunar village design
- Irene Schlacht, specialist on habitability in space and integrated design processes for space application
- Andrew Aldrin, professor in space studies, advisor on space tourism issues and mentor to one of the authors.

All of these specialists were chosen and interviewed depending on their specific field of expertise, which resulted in varying interview questions that followed a semi-structured interview design. The combination of a literature review and interview study enabled a broad and deep familiarization with the problem landscape, leading to a set of key qualitative (as opposed to quantitative) findings. At a deeper level, by employing these two methods in parallel, a foresight exercise [15] was made possible (trend-forecasting [16]), enabling us to find a plausible future, which then served as the starting point for answering the research question that constitutes the main objective of this paper.

Once the mission goal was established, a research question defined, and difficulties of the current situation identified, the "Who, What, Where, When, Why, and How" ("WWWWWH") method [13] was applied. Following a recommendation from specialist Antonio Fortunato, we employed this method to review key aspects of a selected future scenario. Through the WWWWWH method a type of check-list of the most important W-questions is created, which provides an excellent point of reference for the ideation design methods in the settlement conceptualization part of this work.

To step out of the exploratory divergence phase and into the convergence phase, design methods, such as a Persona and User-journey-mapping, were used. These methods are useful to illustrate unknown situations and specific touch-points, while uncovering unknown locations and needs of the users in the process [13]. At the same time, they do not require a real situation, nonetheless they allow a realistic exploration of an abstract, or in this case "futuristic", problem. This makes these two methods particularly suitable for abstract future situations like the one that is the subject of this paper. Furthermore, the observations made by architect Christopher Alexander's *A Pattern Language* (1977) [5] on, for example, the development of a community through buildings and making people feel comfortable in a habitat by specific characteristics found in architecture, had a significant influence on the conception phase as well as the final design.

The product was developed through several iteration cycles, which was subsequently visualized and presented through 3D printed Fused Deposition models (FDM), as well as images and illustrations.

**4. Findings**

In this chapter we present the findings of our literature review and the expert inputs. With this we lay the groundwork for the settlement concept

*4.1 Short to long term/duration missions*
*"The demand for habitability is undergoing changes: Following the initial constraints related to short duration missions, we now have to face the conditions of long duration and long distance missions. As a consequence, habitability is required to support performance." [17]*

*Up to today, habitability of space structures has been low. The reason behind this being the origin of development, which is rooted in military advancement, conceptualization by engineers and the use of developmental processes originally designed for machines and other non-human systems. At that point in time, the exploration program was „utilizing men who were able and willing to sacrifice much more than their comfort or safety: their lives."* [17]. With the increase of mission length and distance from earth, during the developments of the MIR and Skylab, habitability became more prevalent. This was the first time that architects or designers, such as Galina Balashova and Raymond Loewy, were introduced to the field. Balashova strived to give cosmonauts a comfortable, visually stimulating workspace: *"Architecture is all about arranging the spaces, not the equipment! The other engineers in our office never really understood that."* [18], however, a majority of her work was never implemented. Raymond Loewy's work was similarly oriented. He strived to provide astronauts with pleasant and easy-to-handle interior design. However, similarly to Balashova's designs, only a small fraction was implemented. The most important being the window in the wardroom of the Skylab. It was at this point in history, against the backdrop of increasing mission duration, when the benefits of including designers and their methodologies in habitat design became evident. *„Habitability, livability-or whatever name is given to the suitability of the environment for daily living-is, as one National Aeronautics and Space Administration (NASA) designer remarked, „a nebulous term at best," one not usually found in the engineer's vocabulary. Besides factors within the engineer's usual responsibilities, such as the composition and temperature of the atmosphere and the levels of light and noise, habitability also encompasses the ease of keeping house, the convenience of attending to personal hygiene, and the provision for exercise and off-duty relaxation. Experience and intuition both suggested that these factors would become more important as missions grew longer."* [19] and these factors are familiar ones to Designers and Architects alike. With increase in mission duration, Human Factors (HF) were implemented. They are defined *"as the scientific discipline concerned with the understanding of interactions among humans and other elements of a system, and the profession that applies theory, principles, data and methods to design in order to optimize human well-being and overall system performance"* (International Ergonomics Association, cited from Schlacht, 14). In her work, Schlacht outlines the meaning of HF in short vs. long duration missions, where HF develops from „ergonomics" (short term), to „habitability" (long term) [17] and the necessity of incorporation HF in a sustainable manner since *„The current low habitability has repercussions on human performance and reliability, increasing the possibility of human error and, as a consequence, jeopardizing mission success."* [17].

Today we are returning to the lunar surface with our Moon-to-Mars Mission (Artemis), extending flight duration and distance from Earth once more. The lunar surface shall become our future home and launchpad into deep-space. This will be the most distant colony we as humans have ever established, intensifying all previously outlined concerns about habitability. It will be the harshest environment we have ever lived in, without atmosphere, constant radiation and meteorite bombardment as well as varying light and dark phases. This will pose entirely new challenges we can not entirely anticipate yet.

*4.2 Importance of Design for Space Application*
As outlined above, with the upcoming Moon missions, human factors and therefore habitability of lunar habitats will become crucial. Not all challenges that will be encountered can be anticipated yet, which is the ideal environment for designers. To navigate the complexity of the interconnected systems and the fragile human in the middle, is the design discipline's strength: *"Designers explore new and sometimes untrodden territories when they ask: what would be a preferred situation? They discover and define opportunities for innovation and improvement. They ask what would be meaningful and valuable for people in their context and dare to take a stance to steer innovation. Designers challenge the way tasks or problems are framed and formulated when they ask what is problematic about this existing situation? They will keep asking questions until they find the root causes and core values that form the starting point for good design. Designers facilitate and drive the development of design solutions that can realize the preferred situation when they ask: what courses of action will best realize the preferred situation and who should I involve to make it happen? They make ideas and visions tangible and iteratively explore their potential in realizing the desired change. They will keep asking how to realize maximum effect with minimum means for* *people and within the complex systems that we find ourselves in today."* [13].

Current developments in the space sector make for an ideal time to introduce design permanently into the developmental process. Whereas it might be difficult to change established developmental processes in governmental agencies, the private space actors might encourage a more rapid paradigm shift. *"It's now up to the private companies to make the journey and a bit like ancient Rome: It was the state that built the roads but it was the private merchants that brought the riches to the city!"* [20].

A recent example is the introduction of touchscreens in the Crew Dragon Capsule from SpaceX. With a new 350 billion dollar market arising [21] we can observe the ongoing shift to human-centered design: "The decision to go with the touchscreen was not thrust on the astronauts all of a sudden. *'It was on the order of at least 5 or 6 years ago that we went out to SpaceX and evaluated a bunch of different control mechanisms,'* said Hurley (Astronaut). *'They were looking at every which way of flying the vehicle, and ultimately they decided on a touchscreen interface.'"*[22].

Methodologies inherent to Design, such as user-interviews, user-feedback or observations are tools to improve issues in the iterative process of designing. However *„Astronauts are particularly rare users if we compare them to the average of the population."* [17], which makes this approach difficult. Furthermore, when questioned about user-feedback on the ISS, one ESA expert responded: "*Astronauts accept their environment. If you ask them what they didn't like, they won't have an answer because they were taught to accommodate the technology they are tasked to handle. Therefore, if you want user-feedback, ask the commercial visitors of space habitats!"* [20], of which, there are even less. However, as Henry Ford has allegedly once said: *"If I had asked people what they wanted, they would have said faster horses."*. Even with these difficulties, design can navigate the unknown through explorative design methodologies like the persona or User-journey mapping. As McGuirk put it in *Moving to Mars*: *„Designers are experts in mediating complexity, including that of bringing technology and human aspects together to work seamlessly."* [23] even when dealing with intangible factors of the quantitative world of engineering, *„(...) the designer as a multidisciplinary team coordinator can be the mediator between engineers and humanities experts."* [17], advocating for the human in the system's center.

One of the biggest assets of design, and a major reason for why it should be a part of our space endeavors, is its capability of envisioning its „preferred solution" by providing imagery and concepts of how we might want to live together in the future. Even though often overlooked, a concept well communicated, through images or renders for example, can spark inspiration that will ignite efforts to bring about a better future on Earth, the Moon and beyond.

*4.3. Key Takeaways*

Combining the need for suitable habitats with the approach of a designer opened up a fascinating new realm of possibilities for the authors of this work. Applying design to tackle space challenges was successful in generating extraordinary outcomes that can be seen as inspiration for future designers and space enthusiasts and provide an outlook to humanity's greatest achievement. All throughout the research journey, this work was received with positive feedback and interest for the outcome.

Through the employed design methodologies, the authors have been able to observe, however, that isolation (and privacy too) is a major topic that needs to be better addressed in lunar habitats. Looking at the designs of the related works in section 3, most habitats are built for groups of 4-6 people. After extended periods of time, this might become challenging on multiple levels. To interact with other inhabitants physically, one must leave the habitat using a space suit (which is very expensive, it is highly unlikely everyone will have their own) and move across the highly hazardous landscape of the lunar surface. Furthermore, when choosing a master-plan approach, or a generally repetitive pattern of buildings, orientation (which is already difficult on the lunar surface due to lighting and landscape constraints) of settlers could be disturbed putting the settler's health at risk as a result. Generally, as discussed in Christopher Alexander's *A Pattern Language*: *„People need an identifiable spatial unit to belong to. They want to be able to identify the part of the city where they live as distinct from all others."* [5]. Furthermore, *„Isolated buildings are symptoms of a disconnected sick society. It is easiest to understand this at the emotional level. The house, in dreams, most often means the self or person of the dreamer. A town of disconnected buildings, in a dream, would be a picture of society, made up of disconnected,* *isolated, selves. And the real towns which have this form, like dreams, embody just this meaning: they perpetuate the arrogant assumption that people stand alone and exist independently of one another. (…)"* [5]. To bridge the distances between habitats, arcades provided an interesting solution. They shield the inhabitants, do not cut spaces through walkways inside the habitat and interesting spaces of encounter, providing ambiguous spaces, allowing spontaneous encounters: *"No social group - whether a family, a work group, or a school group - can survive without constant informal contact among its members."* [5]. To keep the inhabitants grounded, no building should be higher than 4 stories because *„there is abundant evidence to show that high buildings make people crazy."*, Alexander explains that *„high-rise living takes people away from the ground, and away from the casual, everyday society that occurs on the sidewalks and streets (…), it leaves them alone in their apartments. The decision to go out for some public life becomes formal and awkward; unless there is some specific task which brings people out in the world, the tendency is to stay home, alone. The forced isolation then causes individual breakdowns."* [5]

Generally we have observed that there is a profound neglect of socio-cultural factors in current habitat designs even though evidence suggests their importance is already known. As Imhof states: *"Relationships form the basis of human existence. Compared to 'normal' life on Earth, the relevance of social interaction increases considerably when individuals live under extreme conditions in a harsh environment."* [24]. Through the observations of Alexander, we were able to define the most basic components of a city from a holistic perspective, focussing especially on socio-cultural factors that shape belonging in habitat. With the user-journey map of our persona, we were able to define touchpoints, which in addition to Hassell Studios work on their Martian Habitat [25], resulting in the 7 module types the concept consists of. We then created a map to visualize the different areas of living in a general way. This is purposefully done loosely to leave enough empty „patterns" that can be completed through imagination and eventually take shape flexibly depending on the inhabitants' needs towards the first permanent settlement in the lunar lava tubes.

*„We do not believe that these large patterns, which give so much structure to a town or of a neighborhood, can be created by centralized authority, or by laws, or by master plans. We believe instead that they can emerge gradually and organically, almost of their own accord, if every act of building, large or small, takes on the responsibility for gradually shaping its small corner of the world to make larger patterns appear there."* [5] or as the philosopher Alfred Schutz put it, *"whatever has been expected to occur will never occur as it has been expected"*. Contingencies are bound to happen, especially in the long run. A centrally planned solution could never correctly anticipate all of them. A level of decentralism and bottom-up approach might provide the only viable way of dealing with a dynamic and unpredictable environment in the long term. As this is a complex endeavor, we do not believe a single discipline will suffice to tackle the challenge of how we will live in the future. Trans-disciplinarity will be required when building our future home.

**5. Concept**

Whilst the multidisciplinary nature of settlement design makes it a complex undertaking, a product design lens offers a big-picture perspective, enabling a more holistic design approach. We believe orienting ourselves towards structures (i.e. socio-cultural but also urban and architectural systems) is a supportive measure that will help future settlers to become more familiar with spaces, as if from memory. The practice of extrapolating and projecting current knowledge into future design ideas might likewise be considered a branch of critical design, normally utilized to criticize the present situation. In this case, however, we used the positive aspects of today's towns to generate a general framework of familiarity in the future.

*5.1 The Settlement*

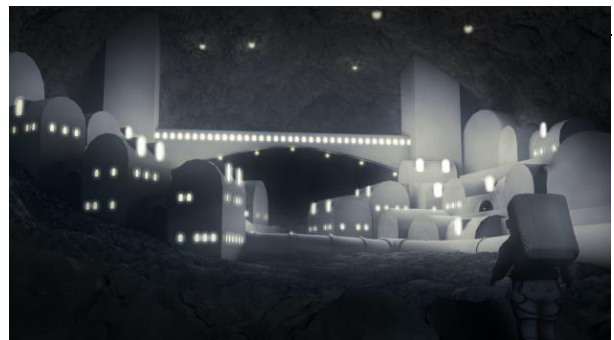

Fig. 1 Render of Village as approached by exploring astronaut, by Tommy Nilsson

The settlement consists of individual buildings, each of which fulfills one or more of the previously identified human needs:
- Private Spaces
- Social Meeting Spaces
- Operational Centers
- Work Spaces
- Storage Spaces
- Agricultural spaces
- Undefined Spaces

All follow a similar pattern language and allow for infinite expansion through the lava tubes of the Moon. Unlike buildings on Earth, the ceilings are considerably higher in order to accommodate movement in the lunar low gravity environment. In addition, none of the buildings is higher than two stories (section 4.3).

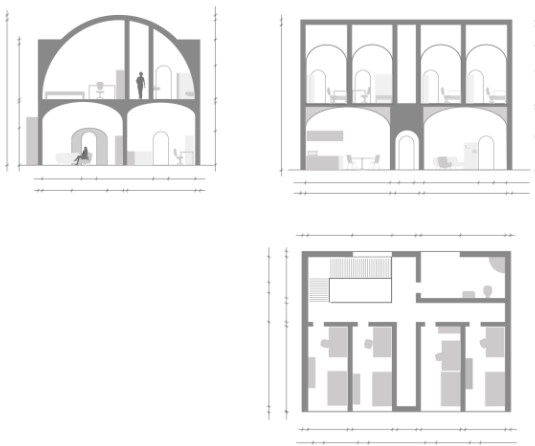

Fig 2. architectural cross-section of Private Quarters. Ceilings are approx 3.5m high due to reduced gravity, Anna Vock

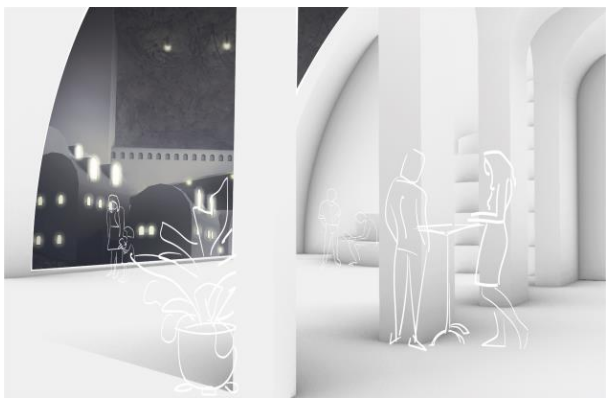

Fig 3. interior view of social spaces occupied by humans enjoying the view onto the city in the background. Pillars and interior arcades provide additional feeling of safety, Anna Vock

Furthermore, spaces in between are connected by arcade structures (Section 4.3) to allow unrestricted walking between buildings and rooms without space suits. All interiors are vaulted to reduce unused space in the corners of the rooms and therefore utilizing space efficiently. It also gives structural integrity, in combination with the pillars. In terms of aesthetics, the selected style is an extension of patterns that can be found across the majority of historic cultures as architectural features.. The large windows provide a view and expand the spaces.

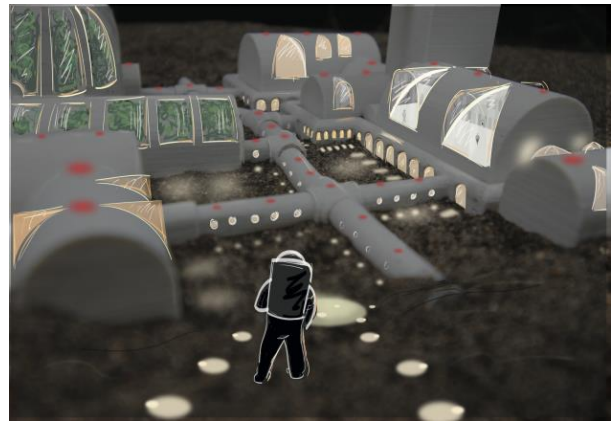

Fig. 4 Visible are the connecting tubes between distances and arcades along the sides of buildings, which shield the inhabitants from vacuum, illustration by Anna Vock

This approach contrasts to the classical "master plan" approach, that has been so frequently adopted by contemporary concepts, thus offering a potential alternative. By leaving settlement patterns semi-defined, residents should be able to shape their own city and thus develop a sense of belonging through exercising a high degree of control over their own living spaces.

It can also be seen from the map that the spaces are not segregated according to their functions. They are meant to be colorfully mixed, much like the cities here on Earth, to foster an urban atmosphere. The result is a mixture of private spaces, offering the residents a place to withdraw to, and social meeting spaces where they come together as a whole or in sub-groups. Undefined spaces can be adapted as needed or provide a neutral lounge or serve as transit spaces. Finally, working spaces can be assigned to focus groups or function as co-working spaces.

The buildings may be repurposed and reused through the use of 3D printed extensions. This fluid transition allows for a dynamic adaptation to the needs of the residents. This enables a "recycling" of spaces as the city continues to grow throughout the cave, constantly evolving hand in hand with the number of inhabitants and their needs. The city has taken on a certain shape in this work, which is solely due to the project framework. As previously explained, this concept is not intended to force buildings into a predetermined master plan; rather, the buildings are to be adapted to the needs of the residents. Only in this way can a settlement grow organically and healthily.

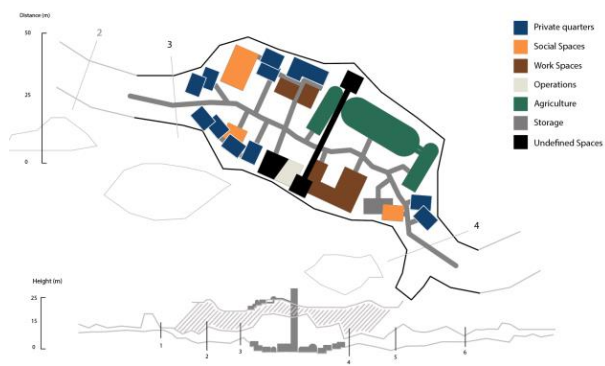

Fig. 5 Map of spaces layed out in a conceptual lava tube structure, by Anna Vock

In Kevin Lynch's *The Image of the City* (1960) empirical research showed the importance of landmarks and reference points for urban space. Lynch outlines a positive effect on the perception of the city [26] by providing the settler with a view over the city while simultaneously functioning as reference points, supporting the mental representation of a perceived space, which could be of much more importance in the monotonous environment of a lunar lava tube (or the lunar surface for that matter) than on Earth.

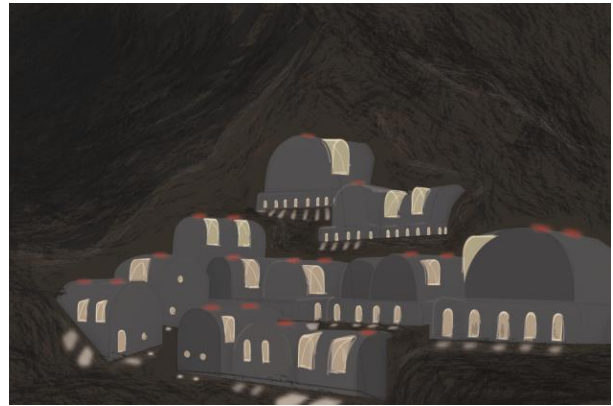

Fig. 6 Conceptional illustration of view from bridge, overlooking the settlement, by Anna Vock

Christopher Alexander also describes the significance of *"high places"* as a vantage point, that one can *"climb up to"* and *"survey your world"* [7]. In addition to serving as a vantage and reference point, the two towers and the bridge stretched between them connect the settlement to the lunar surface.

## 6. Discussion

By virtue of a human-centered approach, the settlement design proposed in this paper focuses on the importance of culture primarily through architecture and urban planning by focussing on establishing social spaces. In doing so, this work shows the importance of a trans-disciplinary treatment of these or similar future habitats and the important role designers play in this endeavor, as mediators of complexity. Thus, this project serves as an applied example of the future tasks of designers all the while offering an outlook on a more human-friendly design of extraterrestrial habitats.

The chosen methods and application sequence proved appropriate and can, in future, be used as a guide for similar projects. Furthermore, by disclosing them and showing their application, professionals outside of design, such as engineers, can become familiar with them and apply them in their own workflow.

### 6.1. Limitations

This work assumes that the habitat will be uniformly 3D printed through in-situ resource utilization. While the technologies for this already exist, they have only been tested on Earth's surface under terrestrial conditions. Therefore, the applicability of the selected production method would also need to be verified prior to implementation. Beyond that, there should be feedback rounds with experts and users to discuss the feasibility of this concept, which unfortunately could not be conducted until this point in time.

*6.2. Future Work*
In the future, it would be interesting to add a virtual reality (VR) phase to the iteration cycles. VR proved very useful in similar situations where user testing of future solutions was necessary. This way, close-to-life data could be collected from users of all backgrounds through observations, questionnaires and interviews. With these results, the product in question could be significantly improved without requiring any real-life prototype development. XR (Extended Reality, umbrella term for Virtual and/or Augmented Reality) offers the possibility to test prototypes in a virtual environment without investing the costs of an analogue prototype. ESA is currently testing the extent to which XR could be applied to spaceflight through various projects, with meaningful results [27]. Therefore, a VR user-study in which users walk through the settlement and give feedback on the architecture and urban planning would be a very interesting prospect for this project. Simultaneously, this could open up a new field of space design which utilizes a shared language (XR) between design and engineering, for the benefit of future space settlers.

**7. Conclusion**
Establishing a lunar outpost will be one of the greatest achievements of humankind. Many years were spent collecting information about humans in space. Now we are entering a new frontier of prolonged presence on the lunar surface. While spaceflight has been engineering dominated, the mission distance and duration will require a greater focus on the human within the system and habitability of interiors. This is the realm of design. Although space design is still in its infancy today, the developments over the past years have demonstrated the importance of drawing on the design practice with its human-centered perspective and bottom-up approach. This, in turn, will ensure that the end users of prospective habitats and their interests will stand firmly in the spotlight as plans for future lunar colonization are being drawn ou This work has provided a methodological approach to such an objective and seeks to give shape to a new concept, that focuses on the importance of belonging and community that can be shaped through architectural and urban planning features in the new environment that awaits us on the lunar surface. Beyond the habitat concept itself, we illustrate the importance of a transdisciplinary approach and common language for this future challenge.

As renowned architect, artist and musician Daniel Libeskind once said: *"It's what the great architects have told us since the beginning of time; that architecture is about culture, it's about what human beings are, where they're going, where they have been, what they want to do. It's not about bricks and mortar and wood only."* [28].

**Acknowledgements**
The authors would like to thank the experts who provided us with their insights and constant feedback on the developments of this paper, as well as the encouragement to dare new approaches.